\documentclass{vietnam}

\def\be{\begin{equation}}
\def\ee{\end{equation}}
\def\bea{\begin{eqnarray}}
\def\eea{\end{eqnarray}}

\def\ga{\mathrel{\mathchoice {\vcenter{\offinterlineskip\halign{\hfil
$\displaystyle##$\hfil\cr>\cr\sim\cr}}}
{\vcenter{\offinterlineskip\halign{\hfil$\textstyle##$\hfil\cr>\cr\sim\cr}}}
{\vcenter{\offinterlineskip\halign{\hfil$\scriptstyle##$\hfil\cr>\cr\sim\cr}}}
{\vcenter{\offinterlineskip\halign{\hfil$\scriptscriptstyle##$\hfil\cr>\cr
\sim\cr}}}}}

\newcommand{\Photo}{}

\begin{document}
\vspace*{4cm}
\title{ULTRA-HIGH ENERGY COSMIC RAYS: RESULTS AND PROSPECTS}

\author{ Karl-Heinz Kampert }

\address{Department of Physics, University of Wuppertal, Germany}

\maketitle\abstracts{
Recent advances in measuring and interpreting cosmic rays from the spectral ankle to the highest energies are briefly reviewed. A knee of heavy primaries and an ankle of light primaries have been observed at about $10^{17}$\,eV. The light component starts to dominate the flux at the ankle in the all particle spectrum at about $4\cdot10^{18}$\,eV and sheds light on the transition from galactic to extragalactic cosmic rays. The prime question at the highest energies is about the origin of the flux suppression observed at $E \ga 4\cdot10^{19}$\,eV. Is this the long awaited GZK-effect or the exhaustion of sources? The key to answering this question is again the still largely unknown mass composition at the highest energies. Data from different observatories don't quite agree and common efforts have been started to settle that question. The high level of isotropy observed even at the highest energies challenges models of a proton dominated composition if extragalactic magnetic fields are on the order of a few nG or less. We will discuss the experimental and theoretical progress in the field and the prospects for the next decade.
}

\section{Introduction}

The advent of the Pierre Auger Observatory and the Telescope-Array (TA) as well as recent data from KASCADE-Grande and IceTop have dramatically advanced our understanding of ultra-high energy cosmic rays (UHECRs). At energies of about $10^{17}$\,eV, a second knee dominated by heavy primaries has been observed~\cite{Apel:2011bx} as well as an ankle-like feature in the energy spectrum of light primaries~\cite{Apel:2013-ankle}. At the highest energies, the GZK-range, the suppression of the flux is now confirmed without any doubt~\cite{Abbasi-08,Abraham-08c}. Moreover, strong limits have been placed on the photon and neutrino components at EeV energies~\cite{Abraham-photon-09,AbuZayyad:2013fu,Abreu:2011tc}, indications are found for the presence of a large-scale anisotropy both below and above the energy of the ankle~\cite{PAbreuetal:2012ve}, and for an anisotropy on smaller angular scales at $E>5.5 \cdot 10^{19}$\,eV~\cite{Abraham-07e,Abreu-10,AbuZayyad:2012th}. 
Particularly exciting -- but also somewhat controversial -- is the observed behavior of the depth of shower maximum with energy which changes in an unexpected, non-trivial way in the data from the Pierre Auger Observatory.  Around $3 \cdot 10^{18}$\,eV it shows a distinct change of $\langle X_{\rm max}\rangle$ with energy and the shower-to-shower variance decreases~\cite{Abraham-xmax-10}.  Interpreted with the leading LHC-tuned shower models, this implies a gradual shift to a heavier composition. The preliminary TA data do not show this trend and are compatible with a proton dominated composition~\cite{Sagawa-2013}. However, the TA data  still suffer from statistics and a change in the cosmic ray (CR) mass composition, such as suggested data from the Pierre Auger Observatory cannot be excluded. Another interesting observation made by both observatories, each taking advantage of hybrid measurements, is a mismatch between energy scales obtained from calorimetric fluorescence observations and particle observations at the ground. In data from the Pierre Auger Observatory, this mismatch can be attributed to a muon deficit in simulations of Extensive Air Showers (EAS) which increases with primary energy and ranges from 30\,\% to about 80\,\% dependent on assumptions about the primary mass~\cite{Farrar-13}.

The increasing mass composition from the ankle towards the highest energies, the high level of isotropies in the arrival direction, and the upper bounds of EeV photon- and neutrino-fluxes raise doubts about the flux suppression observed above $4 \cdot 10^{19}$\,eV being caused (solely) by CR energy losses in the CMB (GZK-effect~\cite{Greisen-66,Zatsepin-66}). Instead, a scenario in which the highest energy CR accelerators reach their limiting energy already below $E/Z \approx 10^{19}$\,eV appears to best describe the bulk of the data. 

In the following, we shall review the most recent experimental data, examine their uncertainties and limitations and discuss the results in comparison to astrophysical scenarios. Conclusions about directions for the near- and mid-term future will be drawn.

\section{The Cosmic Ray Energy Spectrum from the Ankle to the Highest Energies}

\begin{figure}[t,b]
\begin{minipage}[b]{0.52\textwidth}
\centerline{\includegraphics[width=.9\textwidth]{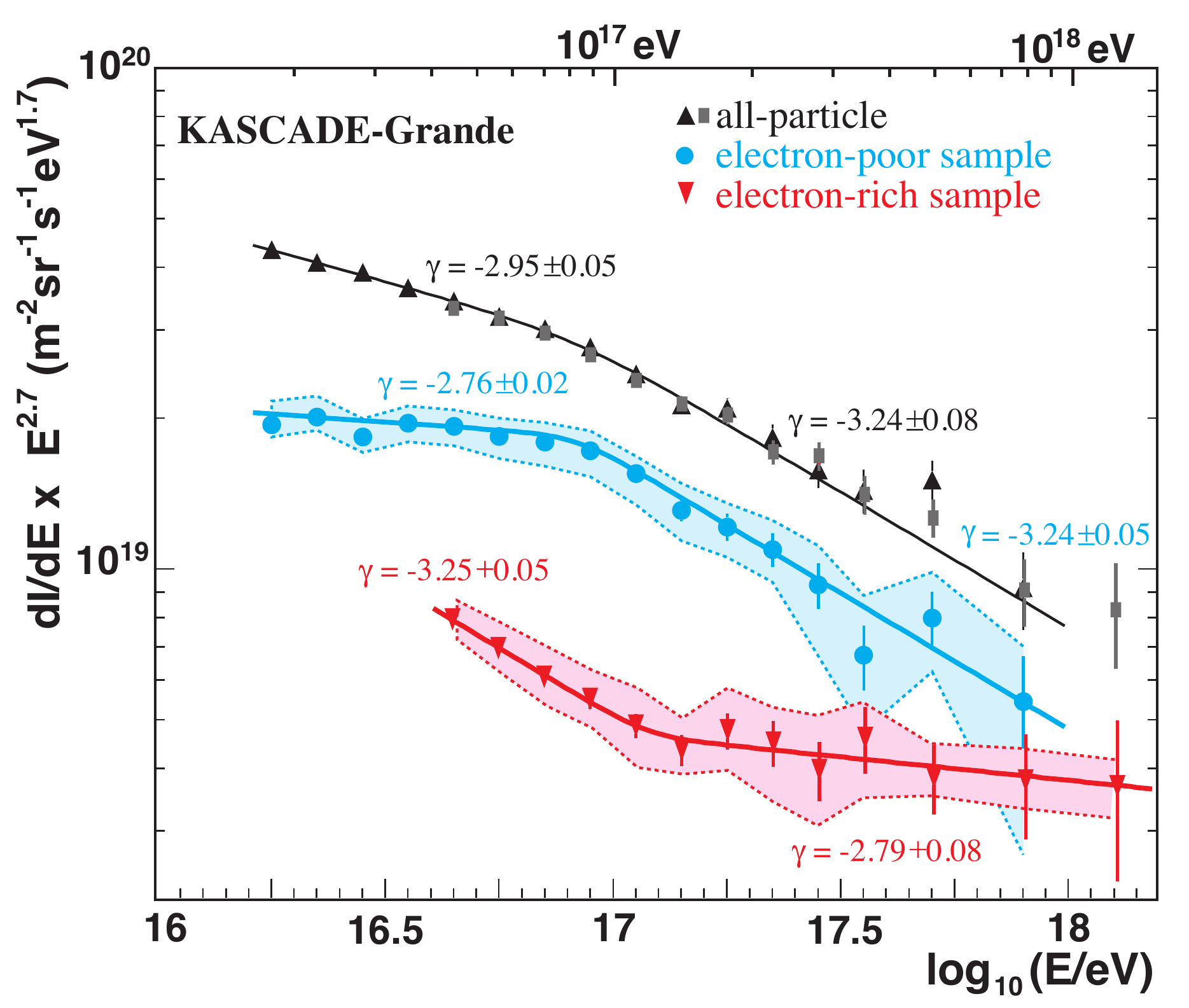}}
\end{minipage}
\begin{minipage}[b]{0.48\textwidth}
\caption[xx]{All-particle, electron-poor, and electron rich energy spectra from KASCADE-Grande. The all-particle (black triangles; 105,000 events) and heavy enriched spectra (blue circles; 52,000 events) are taken from \cite{Apel:2011bx} and the all-particle (grey squares) and light primary spectrum (red triangles; 6,300 events) result from a larger data set with stronger cuts to the light-component applied to select essentially p+He primaries \cite{Apel:2013-ankle}. The bands indicate systematic uncertainties resulting mostly from hadronic interaction models. The heavy enriched data sample exhibits a knee at $10^{16.9}$\,eV with a statistical significance of $3.5\sigma$ while the ankle-like feature in the light component is found at $10^{17.1}$\,eV with a significance of $5.8\sigma$ (Figure from \cite{Kampert-Texas}). \label{fig:kg-spectra}}
\end{minipage} \hfill
\end{figure}

Enormous progress in measurements of CRs has been made recently both in the knee-to-ankle energy range as well as at the highest energies. This has been driven mostly by KASCADE-Grande and IceTop in the knee-to-ankle range \cite{Apel:2011bx} and by Auger and TA in the ankle-to-GZK range. The lesson learned is that, once measured with high precision, the energy spectrum exhibits much more structure and information than just the knee and ankle energies and the indices of an apparent broken power-law like spectrum. The observation of an `Fe-knee' and 'p-ankle' (with ``Fe'' and ``p'' meant as synonym for ``heavy'' and ``light'' primaries, respectively) is a remarkable achievement (c.f.\ Fig.\,\ref{fig:kg-spectra}). The Fe-knee at $8\cdot 10^{16} \approx 26\times3\cdot 10^{15}$\,eV supports the picture of a rigidity scaling in the knee energy range and the p-ankle at $E\simeq 1.2\cdot 10^{17}$\,eV has in fact been expected because of the steep fall-off of the p-component at the knee \cite{KASCADE-05} and the proton dominated composition at the ankle (see next section). Thus, the p-ankle would either mark the transition from Galactic to extragalactic (EG) sources or the onset of a new high energy (Galactic) source population (see e.g.\ \cite{Jokipii:1987uv,Biermann:2012hx,Berezinsky:2014jp}).

\begin{figure}[t,b]
\centerline{
\includegraphics[width=.48\textwidth]{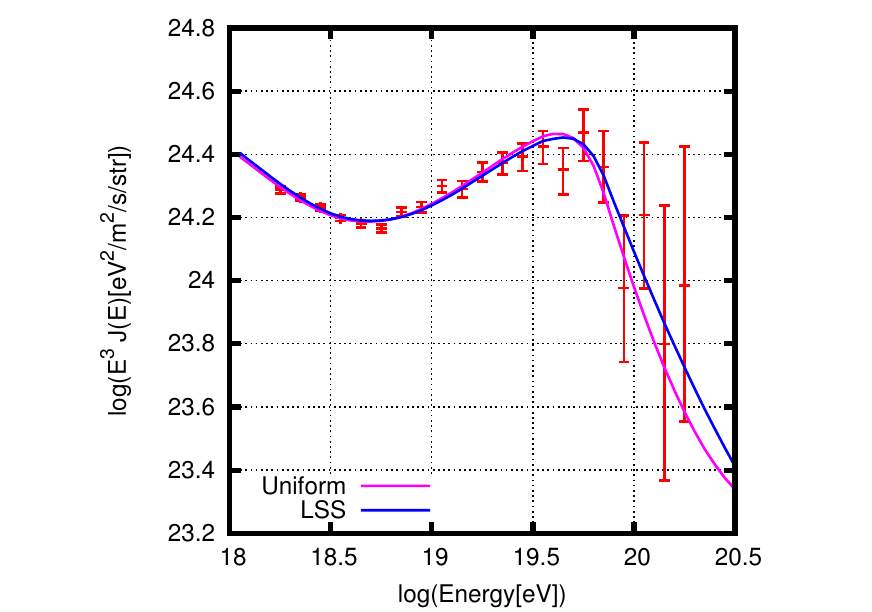}
\includegraphics[width=.48\textwidth]{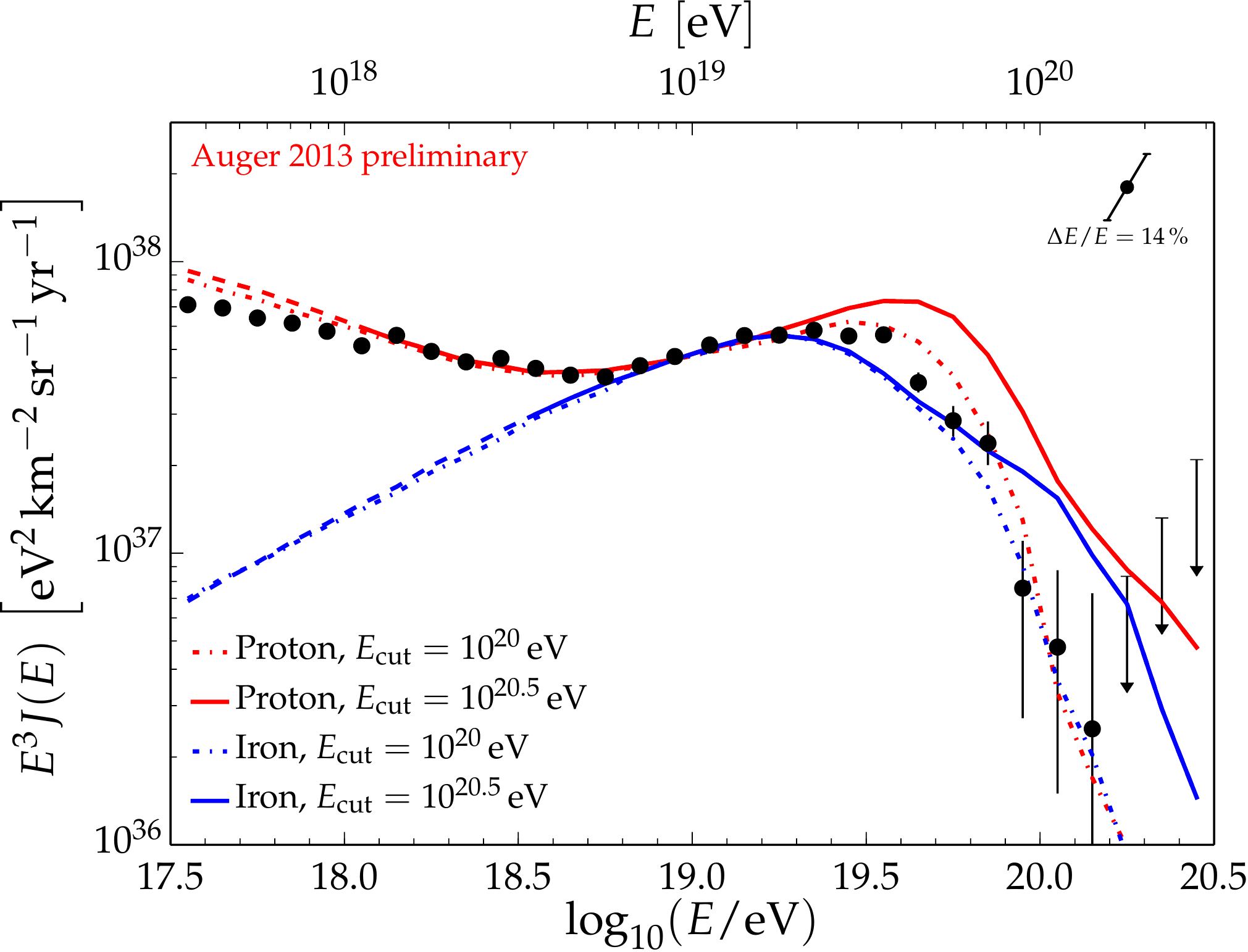}
}
\caption[xx]{Recent measurements of the flux of CRs at the highest energies and comparison to the flux-suppression features expected
in some simple astrophysical scenarios. Left: TA measurements and fits to a model with EG proton sources, assuming a uniform distribution (pink line) or a distribution following the large-scale structure (blue line). The spectral index $\alpha$ and the source
evolution factor (luminosity is assumed to scale as $(1+z)^m$) are best-fitted by $\alpha = 2.36$, $m = 4.5$, and $\alpha = 2.39$, $m = 4.4$ respectively \cite{Kido-13}. Right: Measurements from the Pierre Auger Observatory and comparison to models based on  pure proton composition, with $\alpha = 2.35$ and $m = 5$ (red) or pure iron, with $\alpha = 2.3$ and $m = 0$ (blue) EG sources, with an injection exponential cut-off around $10^{20.5}$\,eV (solid lines) and $10^{20}$\,eV (dotted lines) \cite{Letessier-13}). \label{fig:UHE-spectra}}
\end{figure}

At the highest energies, from the ankle to beyond $10^{20}$\,eV, the Pierre Auger Observatory \cite{auger-04,Abraham-FD-10} is the flagship in the field with an accumulated exposure of about 40\,000\,km$^2$\,sr\,yr at the time of the ICRC 2013 conference. The Telescope Array \cite{AbuZayyad:2012tr}, due to a later start and its more than 4 times smaller area, has collected about a 10th of the events.  A comparison of the energy spectra of the two observatories compared to simplified astrophysical scenarios is presented in Fig.\,\ref{fig:UHE-spectra}. As discussed in great detail in \cite{E-WG}, it is found that the energy spectra determined by the Auger and TA Observatories are consistent in normalization and shape if the uncertainties in the energy scale -- at that time quoted to be about 20\,\% -- are taken into account.  This is quite notable and demonstrates how well the data are understood.  Taking advantage of more precise measurements of the fluorescence yield, of a deeper understanding of the detector and consequently improved event reconstruction, and of a better estimate of the invisible energy, the Pierre Auger Collaboration has recently corrected their energy scale upwards by 10-15\,\% dependent on primary energy. This has reduced the systematic uncertainty of the Auger energy scale to 14\,\% \cite{Verzi-13} and has further reduced the differences between the two experiments.  The spectra in Fig.\,\ref{fig:UHE-spectra} clearly exhibit an ankle at $\sim 4\cdot10^{18}$\,eV and a flux suppression above $\sim 4\cdot10^{19}$\,eV. As mentioned above, the ankle in the all-particle spectrum can be understood by extrapolating the hard p-like and the steep all-particle spectra in Fig.\,\ref{fig:kg-spectra} from the p-ankle towards higher energies. Correspondingly, the composition is expected to change from heavy to light across this energy range. 

The flux suppression at the highest energies is in accordance with the long-awaited GZK-effect \cite{Abbasi-08,Abraham-08c} but appears stronger in the Auger than in TA data. Moreover, as we shall discuss below, the change of the composition seen in the Auger data at the highest energies suggests that the flux suppression is caused  primarily by the maximum acceleration energy of sources or of a source population rather than by CR energy losses in the CMB.

\section{Cosmic Ray Composition and Interaction Models}

Obviously the all-particle energy spectrum by itself, despite the high level of precision reached, does not allow one to conclude about the origin of the spectral structures and thereby about the origin of CRs in different energy regions. Additional key information is obtained from the mass composition of CRs. Unfortunately, the measurement of primary masses is the most difficult task in air shower physics as it relies on comparisons of data to EAS simulations with the latter serving as reference \cite{Kampert:2012hg,Engel-11}. EAS simulations, however, are subject to uncertainties mostly because hadronic interaction models need to be employed at energy ranges much beyond those accessible to man-made particle accelerators. Therefore, the advent of LHC data, particularly those measured in the extreme forward region of the collisions, is of great importance to CR and EAS physics and have been awaited with great interest \cite{Kampert-ISVHECRI12}. Remarkably, interaction models employed in EAS simulations provided a somewhat better prediction of global observables (multiplicities, $p_\perp$-distributions, forward and transverse energy flow, etc.) than typical tunes of HEP models, such as PYTHIA or PHOJET \cite{dEnterria:2011kga}. This demonstrates once more that the CR community has taken great care in extrapolating models to the highest energies. Moreover, as demonstrated e.g.\ in \cite{PAbreuetal:2012vi}, CR data provide important information about particle physics at centre-of-mass energies ten or more times higher than is accessible at LHC. The $pp$-inelastic cross section extracted from data of the Pierre Auger Observatory supports only a modest rise of the inelastic cross section with energy \cite{PAbreuetal:2012vi}.

\begin{figure}
\centerline{
\includegraphics[width=.44\textwidth]{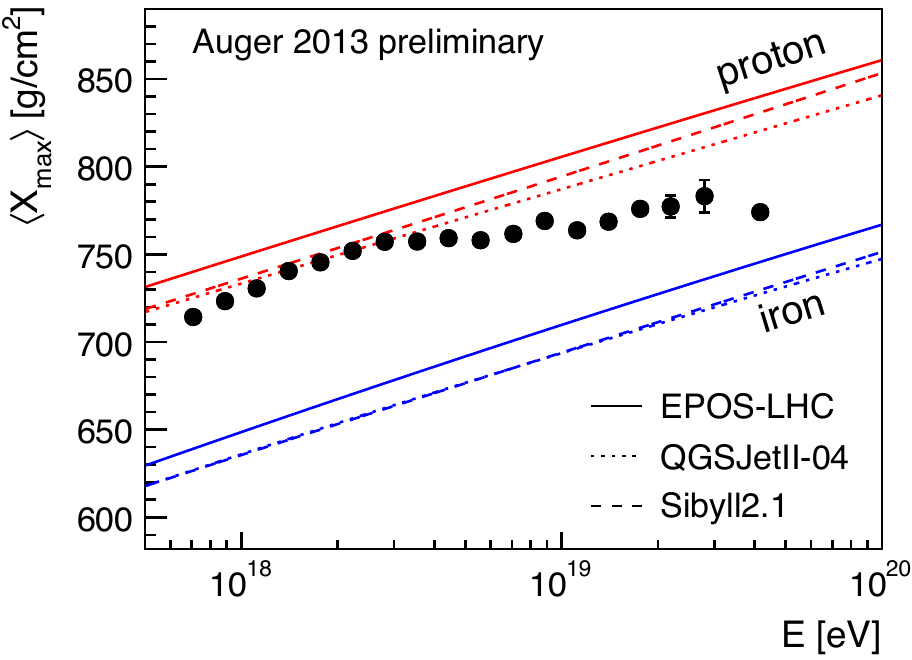}
\includegraphics[width=.44\textwidth]{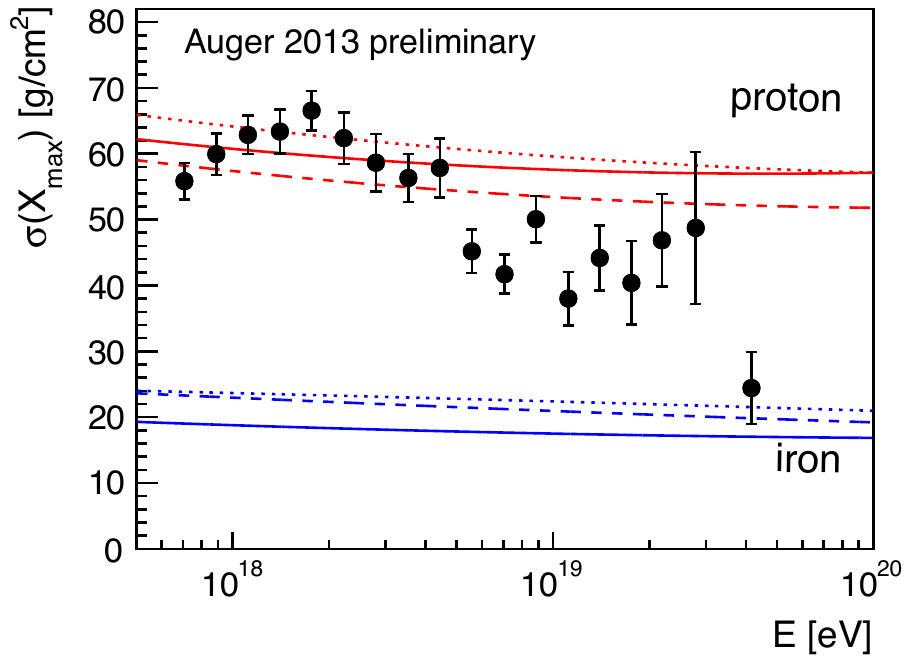}}
\centerline{
\includegraphics[width=.44\textwidth]{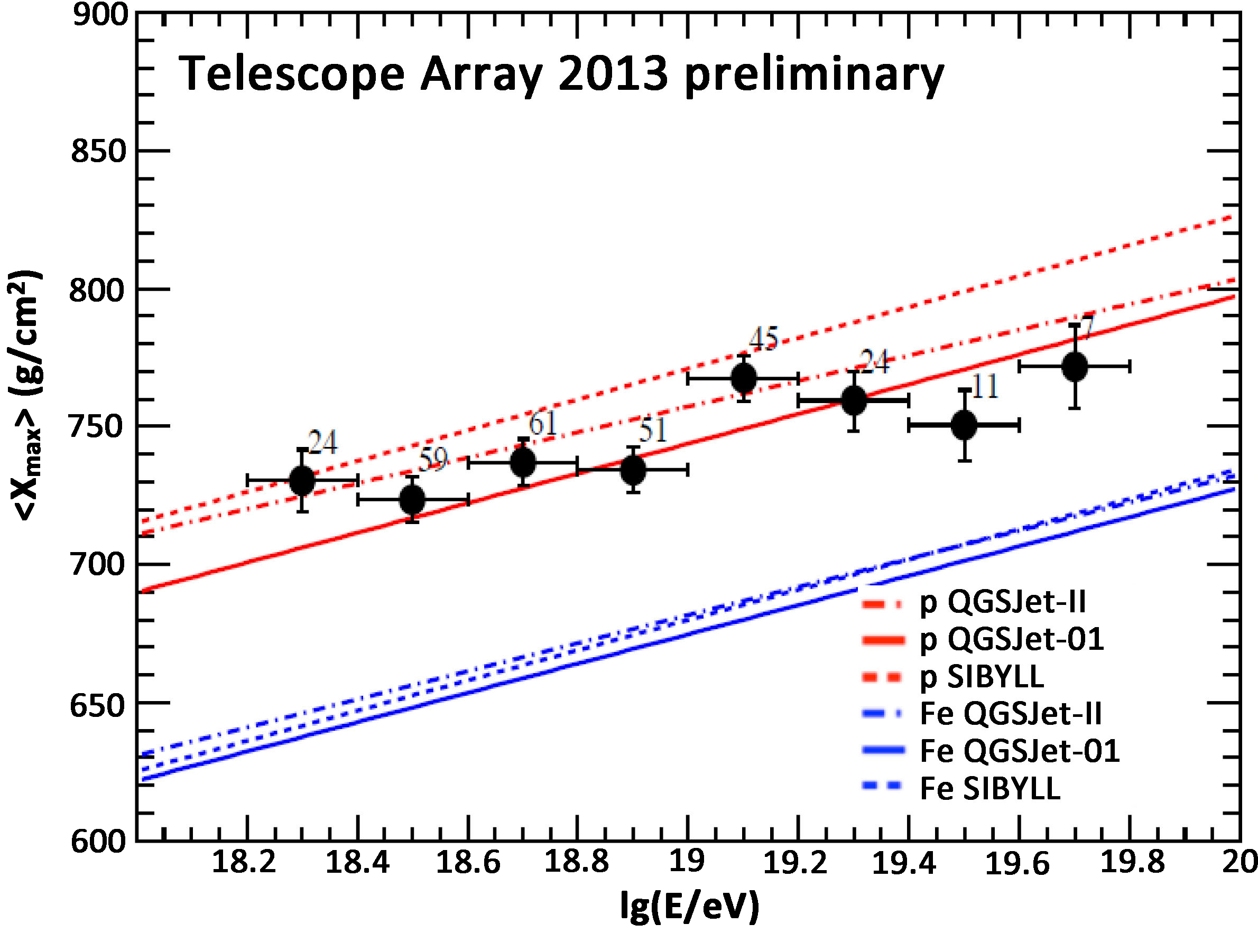}
\includegraphics[width=.44\textwidth]{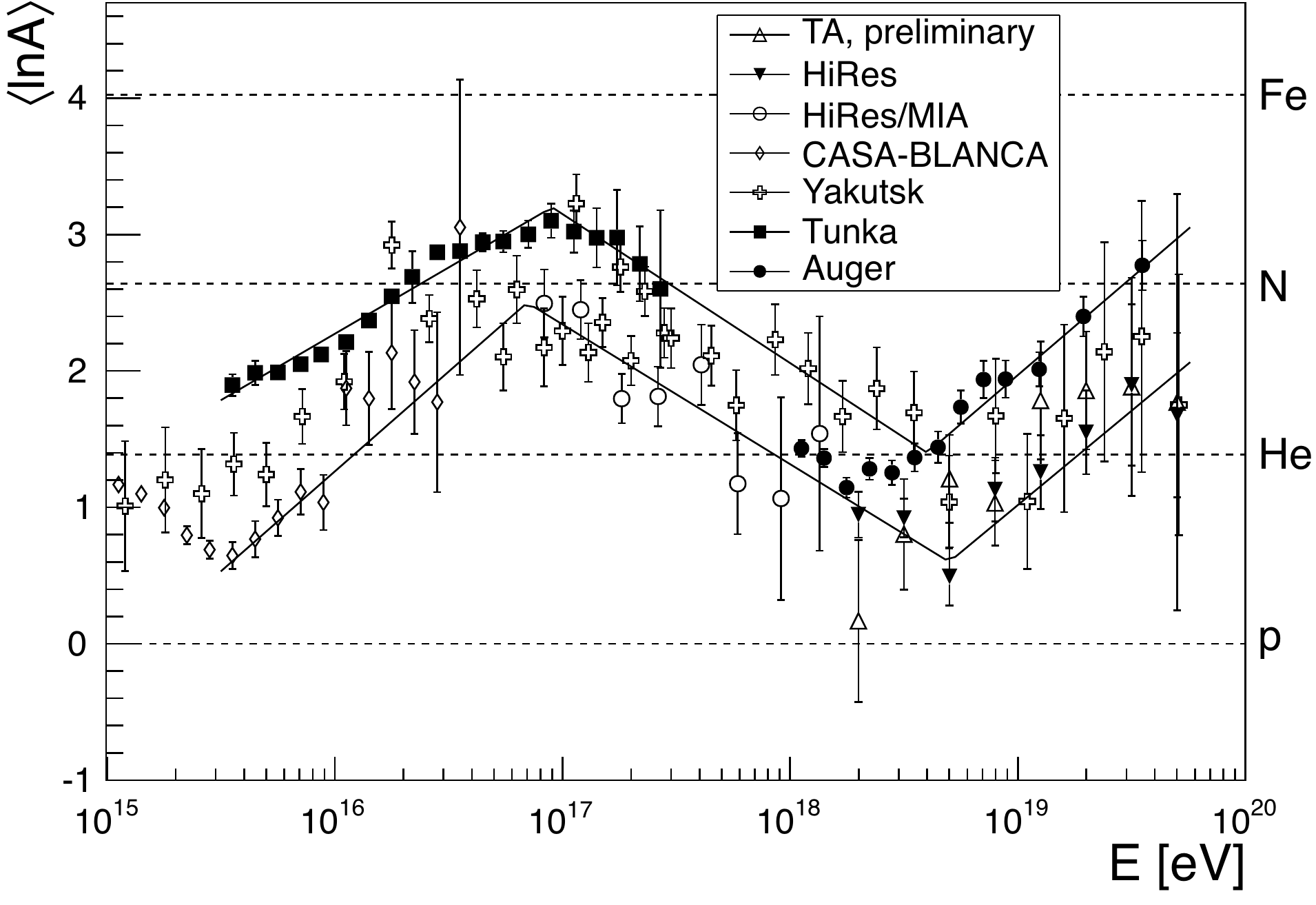}
}
\caption[xx]{Top: Evolution of $\langle X_{\rm max}\rangle$ and $\sigma(X_{\rm max})$ with energy in data from the Pierre Auger Observatory~\cite{Letessier-13}.
Bottom left: $\langle X_{\rm max}\rangle$ as a function of energy from TA~\cite{Sagawa-2013}. Bottom right: Average logarithmic mass of CRs as a function of energy derived from $X_{\rm max}$ measurements with optical detectors for the EPOS 1.99 interaction model. Lines are estimates of the experimental systematics, i.e.\ upper and lower boundaries of the data presented \cite{Kampert:2012hg}.
\label{fig:compos}}
\end{figure}

A careful analysis of composition data from various experiments has been performed and reviewed in~\cite{Kampert:2012hg,compos-WG}. Updated results from the TA and Auger Observatories as well as a comparison of the two were presented at the ICRC 2013 with exemplary results depicted in Fig.\,\ref{fig:compos}. The data from the Pierre Auger Observatory (Fig.\,\ref{fig:compos} top) confirm the earlier results of Auger~\cite{Abraham-xmax-10} and suggest an increasing mass composition above $4\cdot10^{18}$\,eV when compared to post-LHC interaction models. The TA data are compatible with a proton dominated composition at all energies (Fig.\,\ref{fig:compos} bottom left) but have much larger uncertainties and are compared to pre-LHC interaction models. Both collaborations have jointly investigated this difference, e.g.\ by injecting the measured composition from the Pierre Auger Observatory into the TA Monte Carlo. The output of that study shows that, given the present statistics, the proton- and Auger-like composition cannot be distinguished from one another.~\cite{Hanlon-13} It will be interesting to see this puzzle being solved in the near future both by refined and improved reconstruction and analysis techniques, as well as by collecting more data.

A (pre-ICRC 2013) compilation of composition data from various experiments is depicted in (Fig.\,\ref{fig:compos} bottom right). These data complement those of the energy spectrum in a remarkable way. As can be seen, the breaks in the energy spectrum coincide with the turning points of changes in the composition: the mean mass becomes increasingly heavier above the knee, reaches a maximum near the 2$^{\rm nd}$ knee, another minimum at the ankle, before it starts to modestly rise again towards the highest energies. Different interaction models provide the same answer concerning changes in the composition but differ by their absolute values of $\langle \ln A \rangle$~\cite{Kampert:2012hg}.

\begin{figure}
\centerline{\includegraphics[width=.35\textwidth]{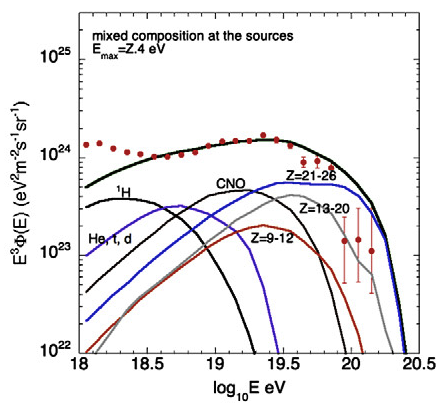}
\includegraphics[width=.44\textwidth]{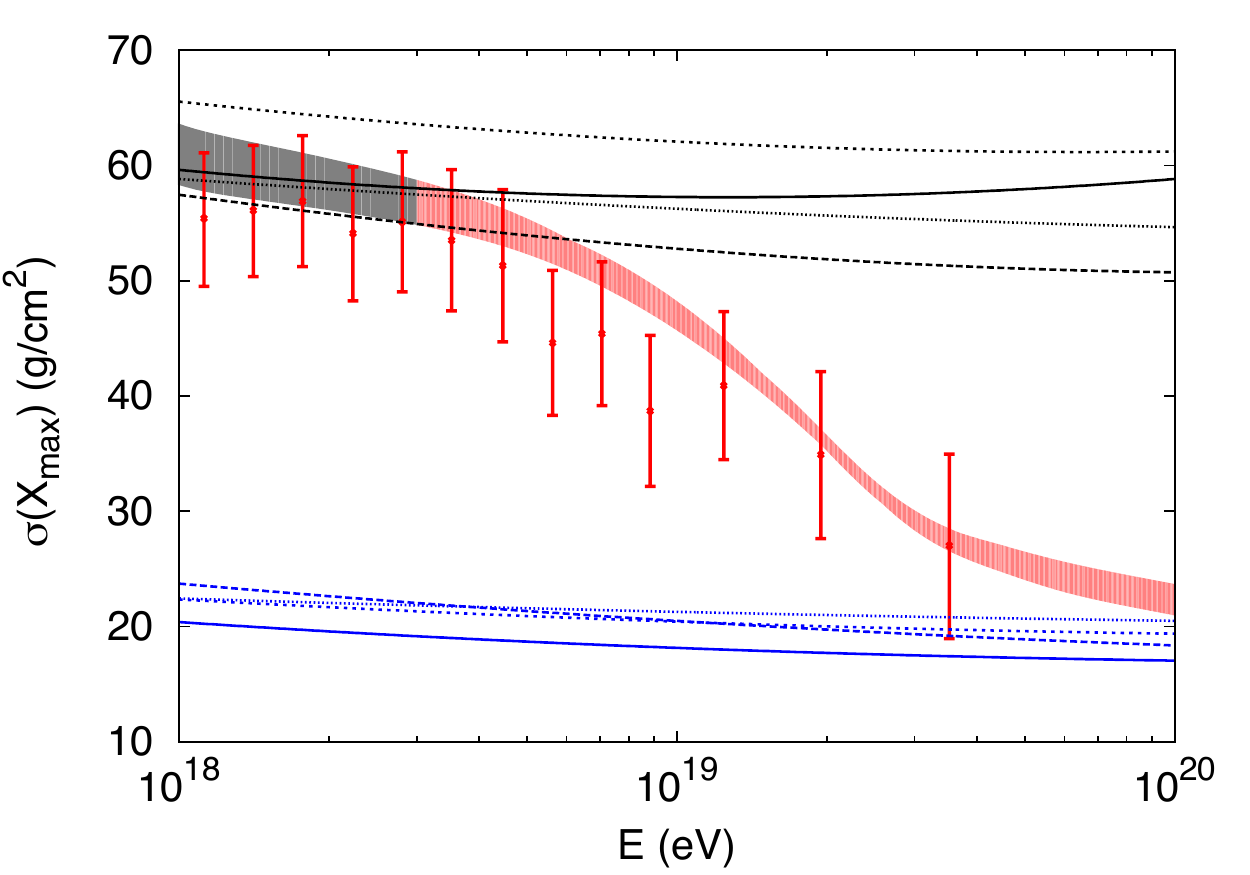}}
\caption[xx]{Left: Propagated CR spectrum assuming a mixed composition similar to the Galactic one with a maximum energy at the sources of $E_{\rm max}(Z) = Z\times4\cdot10^{18}$\,eV and a spectral index $\beta = 1.6$~\cite{Allard:2011ul}. Right: RMS($X_{\rm max}$) from \cite{Blasi:2014ua} assuming $\beta=1.0$ and $E_{\rm max}(Z) = Z\times5\cdot10^{18}$\,eV.\label{fig:allard}}
\end{figure}

The importance of measuring the composition up to the highest energy cannot be overstated as it will be the key to answering the question about the origin of the GZK-like flux suppression and the transition from galactic- to extra-galactic CRs. The same mechanism of limiting source energy that appears to cause the knee and the increasingly heavy mass composition above the knee may work also for EG CRs above the ankle. Thereby, the break at $\sim 4\cdot10^{19}$\,eV may mark the maximum energy of EG CR accelerators, rather than the GZK-effect. This is demonstrated in Fig.\,\ref{fig:allard}, where propagated CR spectra and RMS($X_{\rm max}$) are shown for a maximum energy at the source of $E_{\rm max}(Z) = Z\times4\cdot10^{18}$\,eV and assuming a slightly modified galactic CR mass composition and a hard spectral source index of $\beta = 1.6$~\cite{Allard:2011ul} and 1.0, respectively~\cite{Blasi:2014ua}. Similar results are reported e.g.\ in \cite{Taylor:2014vy}. Clearly, such an -- in view of the hard spectral index -- exotic scenario provides a good description of the energy spectrum. It should be noted, however, that none of the simulations of~\cite{Allard:2011ul,Blasi:2014ua,Taylor:2014vy} account for diffusion of CRs which alters the spectrum below $10^{19}$\,eV so that the true injection spectrum may be steeper, i.e.\ less exotic~\cite{Mollerach:2013ek}. Other than the GZK-like interpretation, the maximum energy scenario describes not only the energy spectrum, but also the observed $\langle X_{\rm max} \rangle$ {\em and} the fluctuation RMS($X_{\rm max}$) from the Pierre Auger Observatory. 

The mixture of light and intermediate/heavy primaries at the highest energies predicted by the maximum-energy model may also explain the low level of directional correlations to nearby AGN. Enhancements, presently foreseen by the Pierre Auger Collaboration will address this issue (see below). Improving the composition measurement in the ankle region will be the key also to discriminate between different models proposed to explain the transition from galactic to EG CRs. This has been a prime motivation for the HEAT and TALE extensions of the Pierre Auger and TA Observatories, respectively \cite{Mathes-11,Thomson-11}.

\section{Anisotropies at Different Energies and Angular Scales}

\begin{figure}
\includegraphics[width=.55\textwidth]{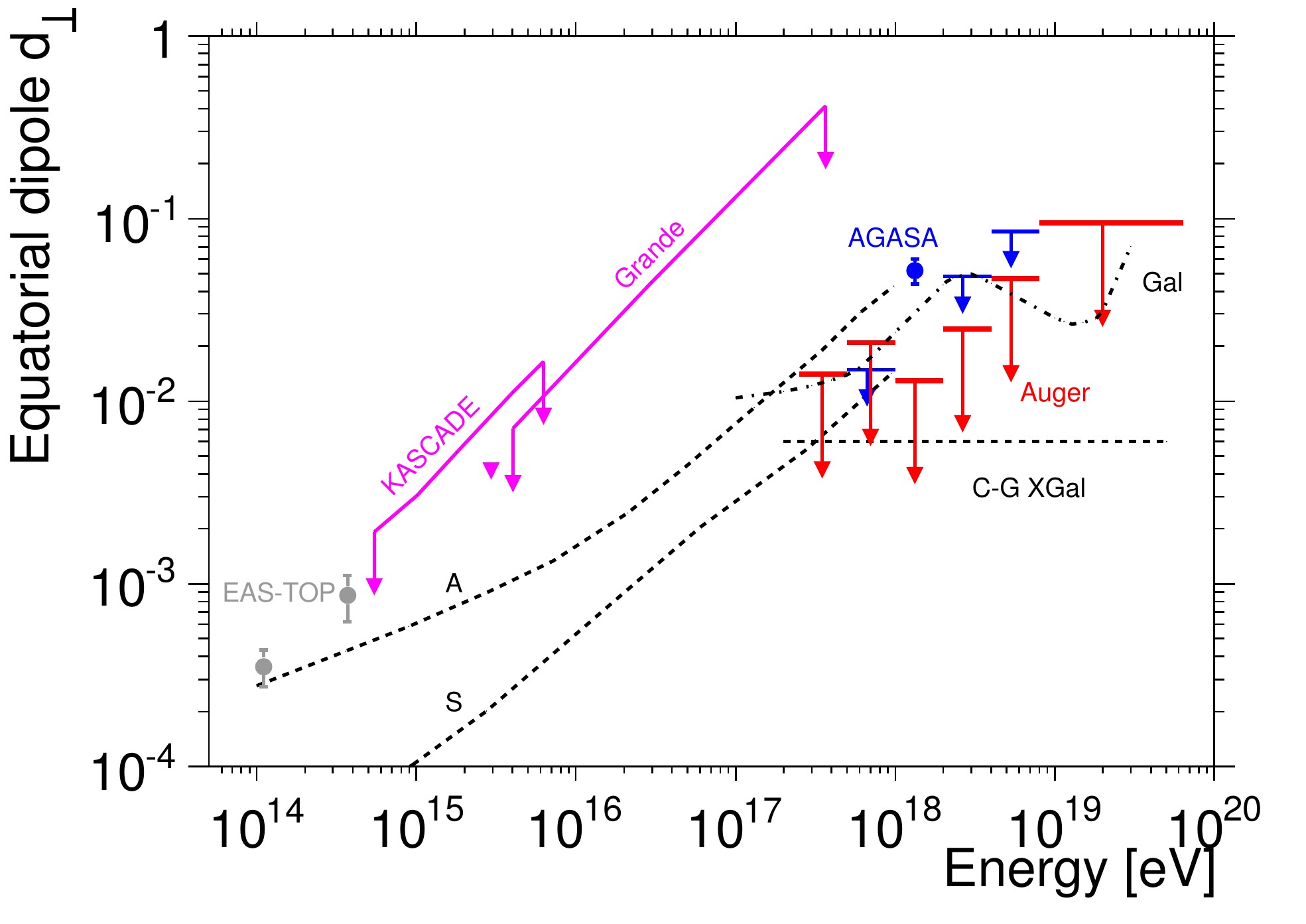}
\includegraphics[width=.45\textwidth]{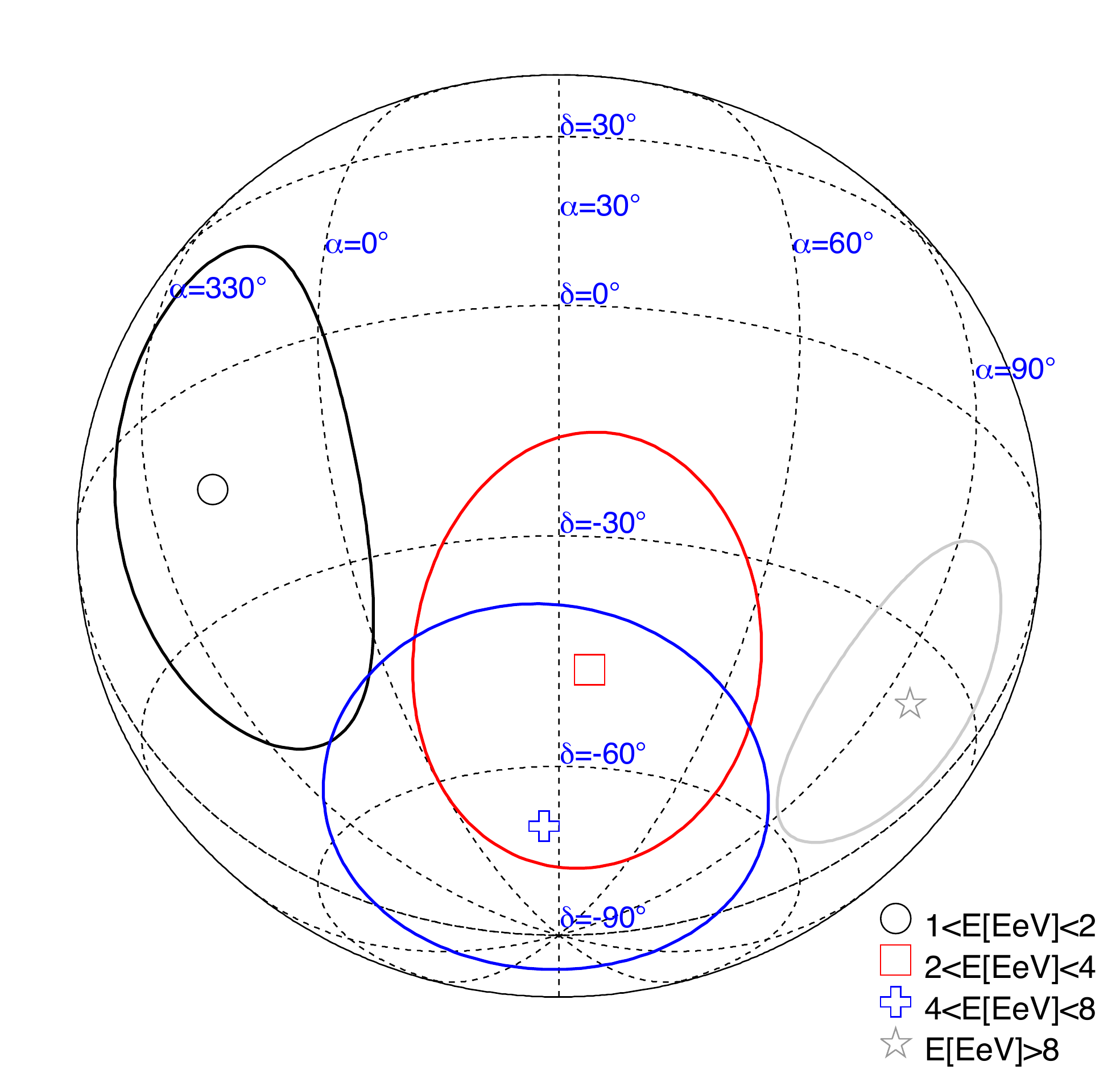}
\caption[xx]{
Left: Upper limits on the equatorial dipole component as a function of energy, from several experiments~\cite{Abreu:2011ve}. Also shown are the predictions up to 1 EeV from two different Galactic magnetic field models with different symmetries (A and S), the predictions for a purely Galactic origin of UHECRs up to a few tens of $10^{19}$\,eV (Gal), and the expectations from the Compton-Getting effect for an EG component isotropic in the CMB rest frame (C-G Xgal) with references given in~\cite{Abreu:2011ve}.
Right: Reconstructed declination and right-ascension of the dipole with corresponding uncertainties, as a function of energy, in orthographic projection~\cite{PAbreuetal:2012ve}.\label{fig:aniso}}
\end{figure}

A long-standing problem of UHECRs is the high level of isotropy observed even at energies beyond $10^{18}$\,eV, i.e.\ into energy regions where CRs cannot be confined by diffuse propagation within our own galaxy. This is illustrated in Fig.\,\ref{fig:aniso} where upper limits of equatorial dipole components start to contradict expectations for CRs of purely galactic origin propagating through different Galactic magnetic field models. However, as pointed out in~\cite{Giacinti:2012hn}, diffusion breaks down for propagation of CRs within our Galaxy at $E \ga 10^{17}$\,eV.
Recently, the Pierre Auger Collaboration reported the first large scale anisotropy searches as a function of both right ascension and declination. Again, within the systematic uncertainties, no significant deviation from isotropy is revealed and the upper limits on dipole and quadrupole amplitudes challenge an origin from stationary galactic sources densely distributed in the galactic disk and emitting predominantly light particles in all directions. In Fig.\,\ref{fig:aniso} (right), the dipole directions are shown in orthographic projection with the associated uncertainties, as a function of the energy. Both angles are expected to be randomly distributed in the case of independent samples whose parent distribution is isotropic. It is thus interesting to note that all reconstructed declinations are in the equatorial southern hemisphere, and to note also the intriguing smooth alignment of the phases in right ascension as a function of the energy.

Directional correlations of the most energetic CRs with nearby AGN observed by the Pierre Auger Observatory provided the first signature about anisotropies of the most energetic CRs and thereby about their EG origin \cite{Abraham-07e,Abreu-10,Abraham-08a}. Initially very strong, the fraction of Auger events above 55 EeV correlating within $3.1^\circ$ with a nearby ($z\le0.018$) AGN from the VCV-catalogue has stabilized at a level of $(33\pm5)$\,\% \cite{Kampert-ICRC}. With an accidental rate for an isotropic distribution of 21\,\%, this corresponds to a chance probability of less than 1\,\%. Recently, TA reported a correlation fraction of 44\,\% at an isotropic fraction of 24\,\% yielding a chance probability of about 2\,\% \cite{AbuZayyad:2012th}. Thus, the data are in perfect agreement with each other yielding a combined chance probability of observing such a correlation at the $10^{-3}$ level. However, more statistics are needed to consolidate the picture and to allow subdividing data sets in bins of related CR observables. The sky region around Cen~A remains populated by a larger number of high energy events compared to the rest of the sky, with the largest departure from isotropy at $24^\circ$ around the center of Cen~A with 19 events observed and 7.6 expected for isotropy, corresponding to a chance probability for this to occur at a level of 4\,\% \cite{Kampert-ICRC}. 

Radio Galaxies remain the most promising candidates for UHECR acceleration. An interesting argument linking UHECR sources to their luminosity at radio frequencies has been put forward by Hardcastle \cite{Hardcastle:2010gm} and he concludes that Radio Galaxies can accelerate protons to the highest observed energies in the lobes if a substantial amount of energy is in the turbulent component of the magnetic field, i.e.\ $B \ga B_{\rm equipart}$, and the Hillas criterion is met. In Cen~A, existing observations do in fact constrain $B \ga B_{\rm equipart}$ for the kpc-scale jet. Moreover, if UHECRs are predominantly protons, then very few sources should contribute to the observed flux. These sources should be easy to identify in the radio and their UHECR spectrum should cut off steeply at the observed highest energies. In contrast, if the mass composition is heavy at the highest energies then many radio galaxies could contribute to the UHECR flux but, due to the much stronger deflection, only the nearby Radio Galaxy Cen~A may be identifiable \cite{Hardcastle:2010gm}. Of course, such a conclusion depends very much on the strength of the EG magnetic fields and the maximum energy reached in the sources, but it demonstrates how much information could be gained by composition enhanced anisotropy measurements.


\section{New Projects and Outlook}

Motivated by the large body of important experimental findings and new insights, the field continues to evolve very dynamically with new projects being planned or existing ones being upgraded.  
In Siberia, the German-Russian project {\bf HiScore} is planned to be constructed at the Tunka site. This project will use open Cherenkov counters for CR measurements around the knee and will be complemented by radio antennas to explore this new detection technology. {\bf HAWK} is being constructed in Mexico. Although its prime goal is the study of the $\gamma$-ray sky above 100~GeV, it will also contribute to measuring CR anisotropies at TeV-energies. {\bf LHAASO}, mostly driven by the Chinese community and much larger and more complex than HAWK, serves the same scientific goals. 

At the highest energies, the {\bf Pierre Auger} and {\bf TA} collaborations prepare for upgrades in performance and size, respectively: Auger aims at improving the mass composition measurement on a shower-by-shower basis and its particle physics capabilities at the highest energies to answer the question about the origin of the flux suppression. TA aims at increasing the surface detector with a 2\,km grid up to 2800\,km$^2$.
Both collaborations have started to join efforts for a Next Generation Ground-based CR Observatory {\bf NGGO}, much larger than existing experiments and aiming at good energy and mass resolution and exploring particle physics aspects at the highest energies. Four proposed and planned space missions constitute the roadmap of the space oriented community: {\bf TUS, JEM-EUSO, KLPVE, and Super-EUSO} aim at contributing step-by-step to establish this challenging field of research. They will reach very large exposures aimed at seeing CR sources, which will be at the expense of energy resolution, composition measurements, and particle physics capabilities. Given the resources of funding available in the next decade or two, it is unlikely that all of the above mentioned projects can be realized. Thus, priority should be given to complementarity rather than on duplication~\footnote{In fact, in December 2013, the Japanese Space Agency JAXA announced discontinuing support for a JEM-EUSO payload of 2 tons. Alternatives are currently being investigated by the collaboration.}.

\section*{Acknowledgments}
\vspace*{-5mm}
{\small Its a pleasure to thank the organizers of the Rencontres du Vietnam for inviting me to participate in this vibrant conference held at the occasion of the inauguration of the International Centre for Interdisciplinary Science Education (ICISE) in the city of Quy Nhon. I am also grateful for many stimulating discussions with colleagues from the KASCADE-Grande, Pierre Auger and TA Collaborations. Financial support by the German Ministry of Research and Education (Grants 05A11PX1 and 05A11PXA) and by the Helmholtz Alliance for Astroparticle Physics (HAP) is gratefully acknowledged.}

\end{document}